\documentclass[a4paper,11pt]{article}
\usepackage{pos}
\usepackage{soul}
\usepackage{orcidlink}

\usepackage{amsthm}
\usepackage{slashed}
\usepackage[english]{babel}
\usepackage[utf8]{inputenc}
\usepackage{graphicx}
\usepackage{color}
\usepackage{amsfonts,amsthm}
\usepackage{bm,bbm}
\usepackage{xcolor}

\newcommand{\eq}[1]{Eq.~(\ref{#1})}

\title{Holography on the lattice: Evidence from 3D supersymmetric Yang--Mills theory}
\ShortTitle{Holography on the lattice: Evidence from 3D supersymmetric Yang--Mills theory}

\author*[a]{Anosh Joseph~\orcidlink{0000-0003-4288-8207}}
\affiliation[a]{National Institute for Theoretical and Computational Sciences, \\ School of Physics, and Mandelstam Institute for Theoretical Physics,\\ University of the Witwatersrand, Johannesburg, Wits 2050, South Africa}
\emailAdd{anosh.joseph@wits.ac.za}

\author[b]{David Schaich~\orcidlink{0000-0002-9826-2951}}
\affiliation[b]{Department of Mathematical Sciences, University of Liverpool, Liverpool L69 7ZL, United Kingdom}
\emailAdd{david.schaich@liverpool.ac.uk}

\FullConference{The 42nd International Symposium on Lattice Field Theory (LATTICE2025)\\
2-8 November 2025\\
Tata Institute of Fundamental Research, Mumbai, India\\}

\abstract{
We present new results from our lattice investigations of maximally supersymmetric Yang--Mills theory in three dimensions, focusing on its nonperturbative phase diagram. Using a lattice formulation that preserves part of the supersymmetry algebra at finite lattice spacing, we study the spatial deconfinement transition, which holography relates to the transition between localized and homogeneous black branes in the dual gravity theory. Our analysis employs $N_L^2 \times N_T$ lattices with $N = 8$ colors in the SU($N$) gauge group, considering $N_T = 8$, $10$ and $12$, in each case with aspect ratios $\alpha = N_L/N_T \leq 3$. The resulting transition temperatures are consistent with the holographic low-temperature, large-$N$ prediction $T_c \propto \alpha^3$, providing further evidence for the gauge--gravity correspondence in this setting.
}

\begin{document}
\maketitle

\section{Introduction}
\label{sec:intro}

Gauge/gravity duality conjectures that certain strongly coupled large-$N$ gauge theories admit equivalent descriptions in terms of weakly coupled string or supergravity theories.
In particular, maximally supersymmetric Yang--Mills (SYM) theories in $(p+1)$ dimensions, obtained by dimensional reduction of ten-dimensional ${\cal N}=1$ SYM, are conjectured to be dual to string theories containing D$p$-branes~\cite{Maldacena:1997re, Itzhaki:1998dd}.
In the combined large-$N$ and strong-coupling limits, properties of the gauge theory are related to the thermodynamics and geometry of the corresponding D$p$-brane supergravity solutions.
The best-known example is the $p = 3$ case, the AdS/CFT correspondence, whose additional conformal symmetry has enabled extensive analytic study.
However, for direct non-perturbative numerical tests of holography, the cases with $p < 3$ are beautiful, since the corresponding gauge theories are super-renormalizable and more amenable to lattice regularization~\cite{Schaich:2022xgy}.

To perform numerical calculations, the gauge theory must be compactified in the spatial directions.
In the continuum, this corresponds to placing the Euclidean theory on a torus with periodic boundary conditions for bosonic fields.
Introducing anti-periodic boundary conditions for fermions along one cycle allows access to thermal physics and richer phase structure.
In the conventional thermodynamic interpretation, this anti-periodic cycle is identified with Euclidean time, and the torus is taken to be rectangular.
However, supersymmetric lattice constructions often employ lattices that are non-cubical and possess enhanced point-group symmetries, which naturally correspond to skewed tori in the continuum limit~\cite{Catterall:2017lub}.
Such skewing obstructs analytic continuation to a Lorentzian signature.
However, it does not prevent comparison with holographic predictions.
One may formulate the dual supergravity theory with a skewed torus as the asymptotic boundary geometry.
In that case, the large-$N$ 't Hooft limit predicts thermodynamic behavior governed by appropriate Euclidean black hole solutions.

In this work, we focus on the $p = 2$ case, namely three-dimensional maximally supersymmetric Yang--Mills theory.
This theory provides a balance between computational feasibility and rich non-perturbative dynamics.
Earlier lattice investigations of this system~\cite{Catterall:2020nmn, Sherletov:2022rnl} used symmetric lattice volumes ($N_x \times N_y \times N_T$ with $N_x = N_y = N_T$) and verified that the bosonic action density agrees with holographic expectations derived from the free energy density of homogeneous black D2-brane solutions.
We are now generalizing these studies to anisotropic volumes $N_L \times N_L \times N_T$ with $N_L \neq N_T$, enabling exploration of the D2--D0 phase transition predicted on the gravity side.
In the gauge theory, this transition is expected to correspond to a spatial deconfinement transition at low temperature in the large-$N$ limit.

The present proceedings report preliminary results on the critical temperature $T_c$ of the spatial deconfinement transition, building on Refs.~\cite{Sherletov:2023udh, Schaich:2025uuo}.
Working on skewed tori with fixed shape, we vary their overall size relative to the dimensionful 't~Hooft coupling $\lambda = N g_{\text{YM}}^2$ and determine $T_c$ for three aspect ratios $\alpha = N_L / N_T \leq 3$.
Our results show encouraging agreement with the holographic prediction $T_c \propto \alpha^3$~\cite{Morita:2014ypa}, providing further quantitative support for gauge/gravity duality in the D2-brane system.

\section{SYM on a skewed torus and the supergravity dual}
\label{sec:continuum}

We consider three-dimensional maximally supersymmetric Yang--Mills theory in Euclidean signature, defined on a three-torus ${\mathbb T}^3$.
We impose anti-periodic boundary conditions for fermions along one cycle, identified with Euclidean time $\tau \sim \tau + \beta$, while periodic boundary conditions are imposed along the remaining two cycles.
Denoting the spatial coordinates by $x_i$ ($i = 1, 2$), the torus is defined by the identifications $\tau \sim \tau + \beta$ (anti-periodic fermions) and $(\tau, x_i) \sim (\tau, x_i) + \vec{L}_{1, 2}$ (periodic fermions). 
If the vectors $\vec{L}_{1, 2}$ were orthogonal to each other and to $\tau$, the torus would be rectangular and $\beta$ would admit the standard thermal interpretation as inverse temperature.
Instead, motivated by supersymmetric lattice constructions, we consider a \emph{skewed} torus geometry.
Although such a geometry does not allow analytic continuation to a real Lorentzian signature, holographic duality remains applicable: the large-$N$, strong-coupling limit of the gauge theory is described by a dual string theory, which reduces to supergravity in the appropriate limit.

\textbf{Supergravity prediction.}
If the spatial directions are non-compact so that $1/\beta$ is a genuine temperature, the translation-invariant black D2-brane solution yields a free energy density
\begin{align}
\frac{f}{N^2 \lambda^3} = - \left( \frac{2^{13} 3^5 \pi^8}{5^{13}} \right)^{1/3} T^{10/3} \approx - 2.49189 \, T^{10/3},
\end{align}
where $T = 1 / (\beta\lambda)$ is the dimensionless temperature.
Compactifying the theory on a torus does not modify this density, so for a rectangular torus, one has $\log Z = - f \, V({\mathbb T}^3)$. 
Because the supergravity solution is translation invariant, the same expression holds for a skewed torus~\cite{Aharony:2005ew}, even though a direct thermal interpretation is absent.

\textbf{Continuum SYM action and scaling.}
The SYM action takes the schematic form
\begin{align}
S_{\text{SYM}} & = S_{\text{Bos}} + S_{\text{Ferm}} \\
S_{\text{Bos}} & = \frac{N}{4 \lambda} \int_{{\mathbb T}^3} d\tau d^2x \, {\rm Tr} \Big( F^2 + 2 \left(D \Phi_I\right)^2 - \left[\Phi_I, \Phi_J \right]^2\Big) \cr
S_{\text{Ferm}} & = \frac{N}{\lambda} \int_{{\mathbb T}^3} d\tau d^2x \, {\rm Tr} \Big( \psi^T \left(\slashed{D} - \left[\Gamma_I \Phi_I, \cdot \right] \right) \psi \Big), \nonumber
\end{align}
where $\Phi_I$ ($I = 3, \ldots, 9$) are seven adjoint scalar fields arising from dimensional reduction of ten-dimensional $\mathcal N = 1$ SYM, and $\Psi$ is the corresponding adjoint fermion.

At large $N$ and sufficiently small $T$, supergravity predicts
\begin{align}
\frac{\langle S_{\text{Bos}} \rangle}{N^2} = - \left( \frac{2^{13} 3^2 \pi^8 T}{5^{13}} \right)^{1/3} \left( \frac{V({\mathbb T}^3)}{\beta^3} \right).
\end{align}
In the opposite limit of small volume ($T \gg 1$), dimensional reduction arguments give $\langle S_{\text{Bos}} \rangle = - 2 N^2$ at large $N$.

\textbf{Skewed torus geometry from the lattice.}
Our lattice construction arises from the dimensional reduction of the four-dimensional theory formulated on the $A_4^*$ lattice, which possesses five symmetric basis vectors.
Reducing one direction produces a three-dimensional SYM on the $A_3^*$ (body-centered cubic) lattice.
Thermal boundary conditions (periodic bosons, anti-periodic fermions) are imposed along the temporal direction, while all other directions are periodic.
In the continuum limit, this corresponds to a skewed three-torus~\cite{Catterall:2020nmn}.

The geometry may be characterized by the dimensionless parameters $r_L = L \lambda$ and $r_T = \beta \lambda$, where $L$ and $\beta$ are the spatial and temporal lengths, respectively.
For this skewed geometry, the dimensionless temperature is~\cite{Catterall:2020nmn}
\begin{equation}
\label{eq:T}
T = \frac{4}{\sqrt{3}} \frac{1}{r_T}.
\end{equation}

\textbf{Spatial deconfinement and holographic scaling.}
Although the systems we consider are thermally deconfined, a non-trivial \emph{spatial} deconfinement transition is expected.
At low temperature (large $r_T$) and large $N$, holography predicts a first-order transition as $r_L$ decreases: a homogeneous black D2-brane phase at large $r_L$ gives way to a phase of localized black holes (D0 branes)~\cite{Morita:2014ypa}.
On the gauge theory side, this transition is diagnosed by the behavior of Wilson lines wrapping the spatial cycles, constructed from the unitary components of the complexified gauge links.

While the precise critical value of $r_L$ is not known analytically, holography predicts a universal temperature dependence~\cite{Morita:2014ypa} $T \propto r_L^{(p-5)/2}$. 
For $p = 2$, this becomes $T \propto r_L^{-3/2}$. 
Expressing this in terms of the aspect ratio $\alpha = r_L/r_T$ and using $T \propto r_T^{-1}$ then gives
\begin{equation}
T_c \propto \alpha^3,
\end{equation}
which provides the key scaling prediction we test non-perturbatively in this work.

\section{Three-dimensional supersymmetric lattice construction}
\label{sec:lattice}

Supersymmetric lattice gauge theories with at least $2^d$ supercharges in $d \geq 2$ dimensions can be formulated using the method of \emph{topological twisting}, in which the supercharges are reorganized into integer-spin representations of a twisted rotation group.
In this construction, at least one nilpotent twisted-scalar supercharge can be preserved exactly at non-zero lattice spacing.
While such a construction is not required for $(0+1)$-dimensional SYM quantum mechanics --- where perturbation theory shows that no relevant supersymmetry-breaking counterterms arise --- it becomes essential in higher dimensions to control fine-tuning and approach the supersymmetric continuum limit efficiently~\cite{Catterall:2009it}.

\textbf{From four to three dimensions.}
The three-dimensional maximally supersymmetric Yang--Mills theory considered here is obtained by classical dimensional reduction of four-dimensional ${\cal N} = 4$ SYM.
The lattice construction of ${\cal N}=4$ SYM discretizes the Marcus (geometric-Langlands) twist of the continuum theory.
The resulting theory preserves a single nilpotent twisted-scalar supersymmetry ${\cal Q}$ exactly at finite lattice spacing, along with U($N$) gauge invariance and a large $S_5$ point-group symmetry inherited from the underlying $A_4^*$ lattice.
These symmetries strongly constrain radiative corrections; in four dimensions, only a small number of logarithmic divergences appear~\cite{Catterall:2013roa}.
Upon dimensional reduction to three dimensions, these divergences vanish, and no fine-tuning is expected to be required to recover the continuum limit.

The reduced theory naturally lives on the $A_3^*$ (body-centered cubic) lattice, whose four basis vectors correspond to vectors drawn from the center of a regular tetrahedron to its vertices.
In practice, we employ the full four-dimensional lattice construction, using publicly available parallel software~\cite{Schaich:2014pda, susy_code}, and set $N_z=1$ to implement dimensional reduction.
The remaining directions are taken to satisfy
\begin{align*}
N_x & = N_y = N_L, &
N_\tau & = N_T,
\end{align*}
with anti-periodic fermion boundary conditions imposed only along the temporal direction.
In the continuum limit, this produces the skewed three-torus geometry discussed in Sec.~\ref{sec:continuum}.

\textbf{Twisted lattice action.}
The lattice action retains the ${\cal Q}$-exact and ${\cal Q}$-closed structure of twisted continuum ${\cal N}=4$ SYM,
\begin{align}
\label{eq:action}
S = \frac{N}{4 \lambda_{\text{lat}}} \sum_n {\rm Tr} \bigg[ & \mathcal Q \left(\chi_{ab}(n) {\cal D}_a^{(+)} {\cal U}_b(n) + \eta(n) \bar{\cal D}_a^{(-)} {\cal U}_a(n) - \frac{1}{2} \eta(n) d(n) \right) \nonumber \\
& -\frac{1}{4} \epsilon_{abcde}\ \chi_{de}(n + \hat{\mu}_a + \hat{\mu}_b + \hat{\mu}_c) \bar{\cal D}_c^{(-)} \chi_{ab}(n)\bigg].
\end{align}
Here, the indices $a, b, \ldots = 1, \dots, 5$ label the five basis vectors of the $A_4^*$ lattice.
The 16 fermionic degrees of freedom are arranged into twisted fields $\eta$, $\psi_a$, and $\chi_{ab} = - \chi_{ba}$ with 1, 5, and 10 components, respectively.
The complexified gauge links ${\cal U}_a$ combine gauge and scalar fields, leading to U($N$) lattice gauge invariance.

The dimensionless lattice parameters are $r_L = N_L \lambda_{\text{lat}}$ and $r_T = N_T \lambda_{\text{lat}}$, where $\lambda_{\text{lat}} = a\lambda$ with lattice spacing $a$.
The corresponding dimensionless temperature is given in Eq. \eqref{eq:T}.
The continuum limit is obtained by taking
\begin{align*}
N_L,\, N_T & \to \infty, &
\lambda_{\text{lat}} & \to 0,
\end{align*}
while keeping $r_L$ and $r_T$ fixed.

\textbf{Soft deformations.}
For numerical stability, we introduce two soft-supersymmetry-breaking deformations.
The first is a single-trace scalar potential that lifts SU($N$) flat directions,
\begin{equation}
\label{eq:single_trace}
S_{\text{soft}} = \frac{N}{4 \lambda_{\text{lat}}} \mu^2 \sum_{n, a} {\rm Tr} \left[ \bigg(\bar{\cal U}_a(n) {\cal U}_a(n) - {\mathbb I}_N \bigg)^2 \right].
\end{equation}

The second deformation ensures genuine Kaluza--Klein dimensional reduction rather than Eguchi--Kawai volume reduction by explicitly breaking the center symmetry in the reduced $z$-direction:
\begin{equation}
\label{eq:center}
S_{\text{center}} = \frac{N}{4 \lambda_{\text{lat}}} \kappa^2 \sum_n {\rm Tr} \left[ \bigg({\cal U}_z(n) - {\mathbb I}_N \bigg)^{\dag} \bigg({\cal U}_z(n) - {\mathbb I}_N \bigg) \right].
\end{equation}
With $N_z = 1$, this term is gauge invariant.
It forces the reduced link to fluctuate around the identity, thereby fully breaking the center symmetry in that direction and ensuring the correct continuum interpretation.

To recover the supersymmetric continuum limit, we scale the deformation parameters as $\mu = \kappa = \zeta \lambda_{\text{lat}}$, so that supersymmetry is restored as $\lambda_{\text{lat}} \to 0$.
In this work, we use $\zeta = 0.5$ or $0.7$ for different aspect ratios $\alpha=r_L/r_T$.

The full lattice action is therefore
\begin{equation}
S_{\text{lattice}} = S + S_{\text{soft}} + S_{\text{center}}.
\end{equation}

\section{Numerical results and comparison with supergravity}
\label{sec:results}

Our numerical calculations employ the rational hybrid Monte Carlo (RHMC) algorithm, implemented within the publicly available parallel software package for lattice supersymmetry~\cite{susy_code, Schaich:2014pda}.
Since Ref.~\cite{Schaich:2014pda}, the code base has been substantially extended, including the improved lattice action of Ref.~\cite{Catterall:2015ira}, support for $p = 0$ matrix models~\cite{Jha:2024rxz}, implementation of the center-breaking deformation in Eq.~\eqref{eq:center}, and ongoing development for three-dimensional $Q = 8$ SYM~\cite{Sherletov:2022rnl}.
All results presented here are obtained using these updated capabilities.

\textbf{Supersymmetry Ward identity.}
Before turning to the phase structure, we verify that supersymmetry-breaking effects remain well under control.
While Eq.~\eqref{eq:action} preserves a scalar supercharge ${\cal Q}$ exactly, this supersymmetry is softly broken by the deformations in Eqs.~\eqref{eq:single_trace} and \eqref{eq:center}, as well as by the thermal boundary conditions.

We monitor these effects through a ${\cal Q}$-Ward identity that fixes the bosonic action to the $\lambda_{\text{lat}}$- and $T$-independent value $s_B = 9N^2 / 2$.

\begin{figure}
  \centering
  \includegraphics[width=0.5\linewidth]{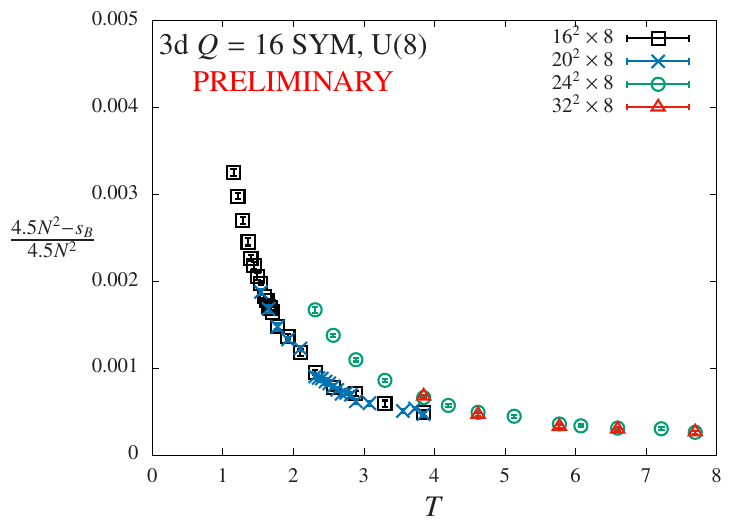}
  \caption{\label{fig:SB}Normalized violations of a ${\cal Q}$-supersymmetry Ward identity involving the bosonic action, $s_B = 9N^2 / 2$, plotted vs.\ the dimensionless temperature from \eq{eq:T}. The violations increase for the stronger 't~Hooft couplings at lower $T$, but remain under half a percent for all the $N_T = 8$ calculations considered here.}
\end{figure}

Figure~\ref{fig:SB} shows normalized violations of this identity as a function of the dimensionless temperature $T$ defined in Eq.~\eqref{eq:T}.
Across the entire parameter range explored here, the violations remain below $0.5\%$ even for our smallest $N_T = 8$.
They increase mildly at lower $T$, corresponding to stronger $\lambda_{\text{lat}}$, indicating that the dominant source of supersymmetry breaking arises from the soft scalar deformations rather than the thermal boundary conditions.
This provides quantitative evidence that our lattice system remains close to the supersymmetric target theory.

\textbf{Thermal deconfinement.}
For a holographic interpretation in terms of black branes to be valid, the gauge theory must remain thermally deconfined.
We verify this by measuring the Polyakov loop (Wilson line in the temporal direction), constructed from the unitary part $U_a$ of the complexified gauge links via polar decomposition ${\cal U}_a = H_a U_a$~\cite{Catterall:2017lub, Catterall:2020nmn}.
The Polyakov loop is normalized so that its maximal value is unity.

\begin{figure}
  \centering
  \includegraphics[width=0.5\linewidth]{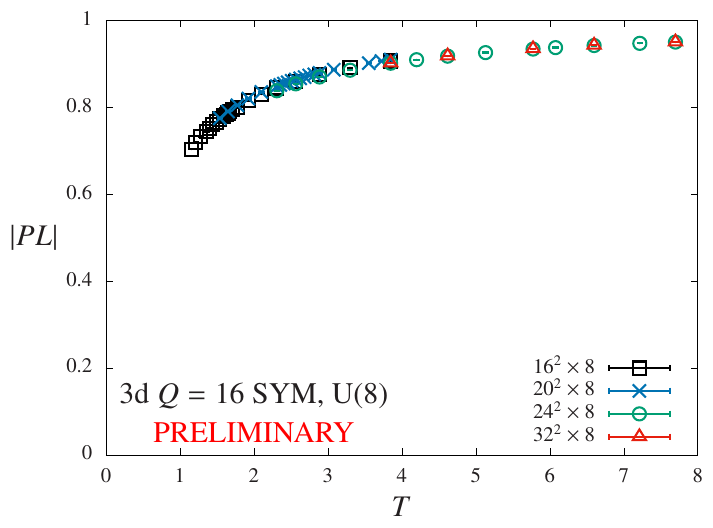}
  \caption{\label{fig:poly}The magnitude of the Polyakov loop vs.~$T$.  For all $N_T = 8$ calculations considered here, $|PL|$ is more than large enough to confirm the thermal deconfinement required for holographic duality to apply.}
\end{figure}

As shown in Fig.~\ref{fig:poly}, we observe $|PL| \gtrsim 0.7$ throughout our parameter range.
This confirms that the system is thermally deconfined for all ensembles considered, ensuring that the comparison with holographic black-brane thermodynamics is meaningful.

\section{Phase diagram results}
\label{sec:phase}

Although the system is always thermally deconfined, holography predicts a non-trivial \emph{spatial} deconfinement transition.
At large $N$ and low temperature (large $r_T$), decreasing $r_L$ should produce a first-order transition from a homogeneous black D2-brane phase to a phase of localized black holes (D0 branes)~\cite{Morita:2014ypa}.
On the gauge-theory side, this transition is diagnosed by the spatial Wilson lines wrapping the $x$ and $y$ cycles.

\textbf{Numerical setup.}
We study lattice volumes $N_L^2 \times N_T$ for three different $N_T = 8$, $10$ and $12$, in each case considering aspect ratios $\alpha = 2$, $2.5$ and $3$.
(For $N_T = 8$ we also consider $32^3\times 8$ lattices with $\alpha = 4$.)
We fix the number of colors to $N=8$.
For each $\alpha$ we vary $\lambda_{\text{lat}}$, thereby scanning diagonally in the $r_T$--$r_L$ plane at fixed $\alpha$.
All results are preliminary, and calculations are ongoing.

\begin{figure}
\centering
\includegraphics[width=0.45\linewidth]{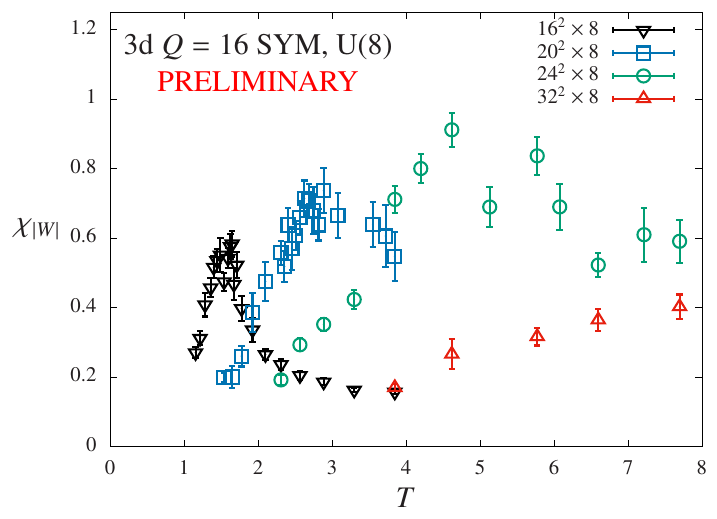}\hfill \includegraphics[width=0.4\linewidth]{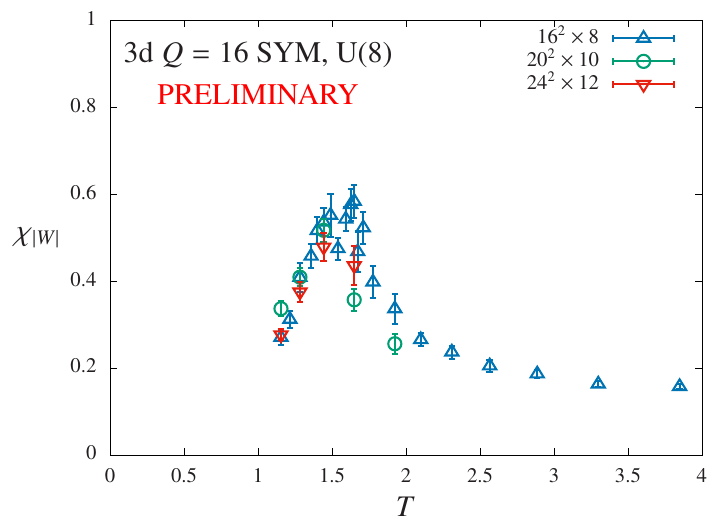}
\caption{\label{fig:signals}Spatial Wilson line susceptibilities vs.\ $T$, considering both different aspect ratios for fixed $N_T = 8$ (left) and different $N_T = 8$, $10$, $12$ for fixed aspect ratio $\alpha = 2$.  Each point combines the Wilson lines in both the $x$- and $y$-directions.  The peaks in the susceptibilities for $\alpha = N_L / N_T \leq 3$ signal the spatial deconfinement transition.}
\end{figure}

\textbf{Signals of spatial deconfinement.}
Figure~\ref{fig:signals} shows the spatial Wilson line susceptibilities, combining the two spatial directions.
In the left plot, we present all four aspect ratios for fixed $N_T = 8$, while in the right plot we consider $\alpha = 2$ for all three $N_T = 8$, $10$, and $12$.
For $\alpha \le 3$, we observe clear susceptibility peaks signaling a transition, with critical temperatures $T_c$ that are independent of $N_T$ within current uncertainties.
From the peaks for all nine volumes we estimate $T_c = 1.54(10)$ for $\alpha=2$, $T_c = 2.6(3)$ for $\alpha=2.5$ and $T_c = 4.4(9)$ for $\alpha=3$.

As $\alpha$ increases, the peaks broaden and statistical fluctuations grow.
For $\alpha \leq 3$, we are increasing statistics and refining the scan in $r_T$ to improve the precision of $T_c$.
For $\alpha = 4$, the transition appears to occur at a significantly higher temperature.
RHMC instabilities become more severe for larger $T$, even after increasing the deformation parameter $\zeta$.
At present, we are unable to reliably reach $T \gtrsim 8$, leaving the $\alpha=4$ transition beyond our accessible range.

\begin{figure}
  \centering
  \includegraphics[width=0.5\linewidth]{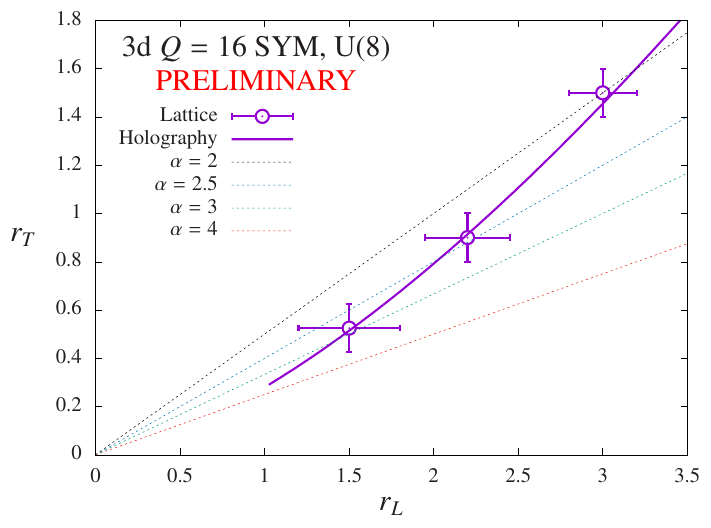}
  \caption{\label{fig:diag}Numerical lattice field theory results for the three-dimensional SYM phase diagram in the $r_T$--$r_L$ plane, compared to the holographic expectation in the form $r_T = 0.28 r_L^{3/2}$ (solid line).  The constant comes from fitting $T = c \alpha^3$, which produces $c = 0.181(10)$.  The dotted diagonal lines show the trajectories we scan with fixed aspect ratio $\alpha = r_L / r_T$.}
\end{figure}

\textbf{Comparison with holography.}
Figure~\ref{fig:diag} summarizes our results in the $r_T$--$r_L$ plane.
We compare the numerical transition points to the holographic expectation written as $r_T = \sqrt{\frac{\sqrt{3} c}{4}} \, r_L^{3/2}$, which corresponds to $T_c = c \alpha^3$.
Fitting these three points produces $c = 0.181(10)$ with $\chi^2/\text{d.o.f.} = 0.97$, giving $r_T \approx 0.28 r_L^{3/2}$.

As a quick consistency check, from this fit we obtain $T_c \approx 1.45$ for $\alpha=2$, $T_c \approx 2.8$ for $\alpha=2.5$, and $T_c \approx 4.9$ for $\alpha=3$.
The $\alpha = 4$ case would correspond to $T_c \gtrsim 11$, consistent with the trend in Fig.~\ref{fig:signals}, though potentially outside the regime where low-temperature supergravity is quantitatively reliable.

Overall, within current uncertainties, our results provide non-perturbative evidence supporting the predicted scaling $T_c \propto \alpha^3$ for the D2--D0 transition.

\section{Conclusions and next steps}
\label{sec:conc}

We have presented preliminary non-perturbative lattice results for the critical temperature of the spatial deconfinement transition in three-dimensional maximally supersymmetric Yang--Mills theory. 
Within current uncertainties, our results exhibit encouraging agreement with the holographic prediction for the D2--D0 transition, in particular the scaling $T_c \propto \alpha^3$ expected from the dual supergravity description of homogeneous black D2 branes and localized black holes. 
These findings provide further quantitative support for gauge/gravity duality in a non-conformal, finite-temperature setting. 
Although the overall picture is now clear, several refinements are underway. 
We are increasing statistics and generating additional ensembles near the susceptibility peaks to improve the determination of $T_c$, especially at higher temperatures where statistical uncertainties grow.
This improved precision will allow non-trivial continuum extrapolations, which correspond to $N_T \to \infty$ with fixed $r_T$.
In addition, we are analyzing the eigenvalue distributions of the spatial Wilson lines, which will allow us to directly confirm the homogeneous and localized character of the two phases~\cite{Catterall:2020nmn}.

An interesting feature of Fig.~\ref{fig:diag} is that uncertainties increase at higher temperatures, in contrast to the behavior observed in analogous two-dimensional studies~\cite{Catterall:2017lub}. 
This difference may be related to the lattice action employed here, motivating future exploration of alternative discretizations.
Another promising direction is the adaptation of gradient flow techniques to three-dimensional SYM, which could significantly reduce statistical noise with modest additional computational cost.
Finally, adding intermediate aspect ratios such as $\alpha = 1.75$ or $2.25$ would provide additional points to test the predicted $T_c \propto \alpha^3$ scaling, though continuum extrapolations in these cases would need to use $N_T = \{8, 12, 16\}$ rather than our current $N_T = \{8, 10, 12\}$.
These improvements should be feasible given currently available computational resources.

A more demanding but important objective is to confirm the first-order nature of the transition predicted by low-temperature holography.
This requires studying the scaling with the number of colors $N$.
While we have explored $N = 4$ and $6$, instabilities become increasingly severe for smaller $N$ with the present action.
Calculations at larger $N$ (e.g.\ $N=10$ or $12$) would be more directly aligned with the large-$N$ limit relevant for classical supergravity, but the computational cost grows faster than $N^3$~\cite{Catterall:2020nmn}, necessitating substantially greater computing resources.

In summary, our results demonstrate steady progress toward controlled large-$N$, strong-coupling tests of holography in three-dimensional maximally supersymmetric Yang--Mills theory.
While larger volumes and larger $N$ will ultimately be required to fully access the classical supergravity regime, the present study already shows clear quantitative consistency with holographic expectations and provides a solid foundation for the next stage of investigations.

\acknowledgments

The work of A.J.\ was supported in part by a Start-up Research Grant from the University of the Witwatersrand. 
D.S.\ is supported by UK Research and Innovation Future Leader Fellowship {MR/X015157/1} as well as Science and Technology Facilities Council (STFC) consolidated grant {ST/X000699/1}.
Numerical calculations were carried out at the University of Liverpool and at the University of Cambridge through the STFC DiRAC facility. \\[8 pt] 

\noindent \textbf{Data Availability Statement:} The data used in this work can be obtained by contacting DS.

\bibliographystyle{JHEP}
\bibliography{lattice25}
\end{document}